# Controllable excitation of quasi-linear and bullet modes in a spin-Hall nano-oscillator

B. Divinskiy,[1*] V. E. Demidov,[1] S. Urazhdin,[2] R. Freeman,[2] A.B. Rinkevich,[3] and S. O. Demokritov[1,3]

[1]*Institute for Applied Physics and Center for Nanotechnology, University of Muenster, Corrensstrasse 2-4, 48149 Muenster, Germany*

[2]*Department of Physics, Emory University, Atlanta, GA 30322, USA*

[3]*Institute of Metal Physics, Ural Division of RAS, Ekaterinburg 620108, Russia*

We experimentally demonstrate that both quasi-linear and nonlinear self-localized bullet modes of magnetization auto-oscillation can be excited by dc current in the nano-gap spin Hall nano-oscillator, by utilizing the geometry with an extended gap. The quasi-linear mode is stable at low driving currents, while the bullet mode is additionally excited at larger currents, and becomes increasingly dominant with increasing current. Time-resolved measurements show that the formation of the bullet mode is delayed relative to the quasi-linear mode by up to 100 nanoseconds, demonstrating that the mechanisms of the formation of these modes are fundamentally different. We discuss the relationship between the observed behaviors and the formation of an unstable nonlinear magnon condensate.

The possibility to generate pure spin currents by the spin-Hall effect (SHE)[1-4] provides unprecedented opportunities for the development of spintronic and magnonic devices. In recent years, it was demonstrated that SHE can be utilized to implement efficient magnetic nano-oscillators – the spin-Hall nano-oscillators (SHNOs)[5,6] – that can serve as nanoscale sources of microwave signals[7-13] and propagating spin waves.[14] These devices are characterized by moderate heat generation, high oscillation coherence, and simple layout. Moreover, pure spin currents produced by SHE were recently shown to enable current-induced excitation of both coherent localized magnetization dynamics, and propagating spin waves in magnetic insulators.[15-18]

The ability of SHE to generate spin currents in a simple thin-film geometry has facilitated the development of a variety of SHNO configurations.[5,6] Two fundamentally different auto-oscillation modes were observed in SHNO, depending on the geometry and the experimental conditions. The quasi-linear mode continuously evolves from the linear eigenmodes of the magnetic system. In contrast, the self-localized bullet mode does not evolve from the linear spectrum, but is instead abruptly spontaneously formed at the auto-oscillation onset.[19] Only one of these modes is typically dominant in SHNO demonstrated so far, even though the other mode may appear under special conditions whose significance is not yet well understood. For instance, the SHNOs formed by a nano-constriction in the ferromagnet/spin Hall material bilayer[9] typically exhibit the quasi-linear auto-oscillation mode,[12,13,20] while the bullet mode has been observed only at low bias magnetic fields and large currents.[21] Meanwhile, SHNOs formed by a nano-gap between sharp electrodes on top of the extended ferromagnet/spin Hall bilayer[22] exhibit bullet-mode auto-oscillations,[23-25] while the quasi-linear mode was found in this geometry only at cryogenic temperatures.[8] In contrast, the recently demonstrated SHNOs based on magnetic insulators,[15] whose geometry is close to that of the nano-gap devices, were shown to exhibit only the quasi-linear auto-oscillation mode.[16] Since the two modes exhibit substantially different oscillation characteristics, beneficial for different specific applications, it is highly desirable to identify the mechanisms controlling the preferential formation of each of these modes, and the approaches to controlling them.

In this Letter, we show that by extending the gap of the nano-gap SHNO in the direction transverse to the current flow, on can achieve controllable generation of the quasi-linear and the bullet modes in the same device. While both modes appear in the same range of driving currents, they exhibit substantially different behaviors at their auto-oscillation onset, clearly indicating that their formation mechanisms are fundamentally different. By analyzing the time-domain evolution of these modes, we find that the emergence of the bullet mode is significantly delayed relative to the quasi-linear mode. The delay can be controlled over a broad range by varying the driving current, which can be utilized to control the auto-oscillation regimes in the time domain. The observed behaviors also provide insight into the mechanisms underlying the formation and the stability of nonlinear magnon condensates.

Figure 1 shows the layout of the test devices. They are based on a bilayer formed by a 4 nm-thick Pt and a 5 nm-thick Permalloy (Py) film, patterned into a square with the side of 5 µm. Two 80 nm-thick Au electrodes separated by a 250 nm-wide gap are fabricated on top of the bilayer. The electrodes are patterned into a trapezoidal shape converging to a 1.8 µm-wide base at the gap. As we shall see below, this geometry enables quasi-linear oscillation dynamics, which is generally suppressed in the previously studied nano-gap SHNOs based on sharp-pointed triangular electrodes, due to the preferential formation of the bullet mode when the size of the active area is comparable to the natural dimensions of the self-localized bullet.[8,22-25]

The auto-oscillations in Py are excited by a dc current $I$ applied between the electrodes. Because of the large difference between the sheet resistances of the electrodes and the Py/Pt bilayer, the electric current in the Py/Pt bilayer is confined predominantly to the region in the gap between the electrodes (inset in Fig. 1). The spin current generated due to the SHE in Pt,[1-4] is injected into the Py layer through the Py/Pt interface. The spin current exerts an anti-damping spin transfer torque on the magnetization *M* of the Py layer (Ref. 26), which is maximized when the static magnetic field $H_0$ is applied perpendicular to the direction of the electric current, as shown in Fig 1. At sufficiently large *I*, the damping in Py becomes completely compensated, leading to the excitation of steady-state magnetization auto-oscillations in the gap.[6]

We detect the SHE-induced magnetization dynamics by using the micro-focus Brillouin light scattering (BLS) spectroscopy.[27] We focus the probing laser light with the wavelength of 532 nm into a diffraction-limited spot on the surface of the Py layer (Fig. 1), and analyze the spectrum of light inelastically scattered from the magnetization oscillations. The detected signal – the BLS intensity – is proportional to the intensity of the oscillations at the selected frequency, at the point of observation. The technique provides information about the magnetization dynamics with simultaneous spectral, spatial, and temporal resolution. To enable BLS measurements in the time domain, the dc current is applied in 200 ns-long pulses with the repetition period of 1 μs.

Figures 2(a) and 2(b) show the BLS spectra recorded at different values of current *I*, with the probing spot positioned in the center of the gap between the electrodes. A narrow intense peak appears in the BLS spectra at *I*≈30 mA, indicating the onset of SHE-induced auto-oscillations (Fig. 2(a)). The central frequency of this peak is close to the frequency $f_{FMR}$ of the uniform ferromagnetic resonance (FMR) in the Py film, determined from independent measurements of thermal magnetization fluctuations. This peak grows with increasing current above 30 mA, while its frequency slightly decreases due to the nonlinear frequency shift.[28] A second peak with the frequency far below $f_{FMR}$ appears in the spectrum at *I*≈33 mA (Fig. 2(b)), indicating a transition to the two-mode auto-oscillation regime. At *I*>33 mA, the intensities of both the high-frequency (HF) and the low-frequency (LF) peaks increase, and the LF peak starts to dominate at large currents.

Figure 3(a) shows the dependences of the intensities of the HF and the LF modes on current. As the current is increased, the intensity of the HF mode continuously evolves from the fluctuation background, while the intensity of the LF is significant even at its abrupt onset at $I$=33 mA. Following the established terminology,[29] these behaviors can be classified as the "soft" and the "hard" onset of auto-oscillations, respectively. The quasi-linear auto-oscillation is generally distinguished by the soft onset, as observed in our experiment for the HF mode, while the nonlinear spin-wave bullet is distinguished by the hard onset observed for the LF mode.[6,19]

This interpretation of the two modes is supported by the current dependences of the auto-oscillation frequencies (Fig. 3(b)). At the onset, the central frequency of the peak corresponding to the HF mode is very close to $f_{FMR}$. This seems to suggest that the HF mode evolves from the linear FMR mode. However, it was recently shown[30] that instead, the spin current injection results in the accumulation of magnons in the lowest-frequency state, reminiscent of the Bose-Einstein condensation of magnons.[31] In the studied 5 nm-thick Py film, the lowest magnon frequency is only about 10 MHz lower than the FMR frequency. Such a small spectral separation is below the resolution of the BLS technique, making it impossible to experimentally distinguish between the two possibilities. Nevertheless, the results of Ref. 30 suggest that the HF mode likely evolves from the lowest-frequency magnon mode.

In contrast to the HF mode, even at the onset of the LF mode its frequency is about 0.6 GHz below the FMR frequency. This result unambiguously demonstrates that the LF mode does not have a counterpart in the linear spectrum, but is formed spontaneously at the auto-oscillation onset, which is one of the essential characteristics of the nonlinear bullet mode.[19]

To gain further insight into the nature the observed modes, we performed spatially resolved BLS measurements of the dynamic magnetization, by scanning the probing laser spot along the nano-gap. Figure 3(c) shows the normalized one-dimensional spatial profiles of the dynamic magnetization, recorded at the frequencies of the LF and the HF modes, at $I$=34 mA. Both auto-oscillation modes are localized in the nano-gap region, consistent with the data in Fig. 3(b), which show that at finite currents the frequencies of the auto-oscillation modes are always smaller than the lowest magnon frequency in the surrounding extended Py film, preventing radiation of propagating spin waves away from the gap. We note that the LF mode is noticeably more localized than the HF mode. This can be attributed to the nonlinear self-localization of the bullet mode, reducing its dimensions below the size of spin current injection region.[19,32] We note that the spatial characteristics of the stable bullet mode are generally determined by the nonlinear properties of the medium, and are expected to be almost independent of the size of spin current injection region. Based on the results of Ref. 32, one expects that the bullet mode should shrink to dimensions below 100 nm, inconsistent with the experimental profile in Fig. 3(c). This discrepancy indicates that the bullet mode in the studied devices with extended gap likely becomes unstable before it is fully formed.

The qualitative differences between the two modes are further elucidated by the time-domain evolution of the current-induced dynamics. Figure 4(a) shows the BLS spectra corresponding to different delay times $t$ relative to the start of the pulse of the driving current $I$=34 mA. At $t$=25 ns, only the HF mode is present in the spectrum. The LF mode emerges at longer delays, suggesting that the mechanism responsible for the formation of this mode is substantially different. This is further illustrated by Fig. 4(b), which shows the time dependences of the intensities of the two modes on the log-linear scale. The intensities of both modes increase with time exponentially, at a similar rate. However, the HF mode emerges from the fluctuation background starting at $t$=0, immediately after the onset of the current pulse, while the LF mode starts to emerge with a significant delay $\Delta t$ of about 27 ns. The observation that the LF mode requires a certain time for its formation indicates that the nonlinear self-localization is facilitated by the initial increase of the dynamical amplitude of the quasi-linear mode. Indeed, for the spin wave bullet to be formed, the dynamical magnetization amplitude must increase to a certain critical level necessary for the onset of nonlinear self-localization.[19]

Figure 4(c) shows the current dependence of the time $\Delta t$ required for the bullet formation. As the driving current $I$ is increased from 33 mA to 36 mA, the value of $\Delta t$ monotonically decreases from 100 ns to about 10 ns. Such a strong dependence can be utilized to develop new types of spin-based devices. For instance, the current pulse duration can be utilized to control the frequency of the microwave output of SHNO.

Finally, we discuss the regime in which the two auto-oscillation modes co-exist. According to the previous studies, the bullet and the quasi-linear mode compete for the same source of the angular momentum provided by the spin current, and are thus mutually exclusive, unless they are spatially separated.[33] Therefore, the simultaneous presence of two spectral peaks in our experiments likely indicates random hopping of the SHNO between these modes.[21,34] This random switching cannot be observed in the temporal dependences in Fig. 4(b), obtained by averaging over multiple pulses of the driving current. These behaviors can be also interpreted in terms of unstable magnon condensation. Because of the attractive magnon-magnon interaction, the initial condensation of magnons at the point of phase space corresponding to the lowest-frequency magnon state – the formation of the coherent quasi-linear HF mode – is followed at sufficiently large magnon densities by the spatial collapse of the condensate, resulting in the formation of the LF bullet mode corresponding to the condensation in the real space. The latter also eventually collapses because of the significant mismatch between the size of the active area and the natural size of the self-localized bullet. We emphasize that, as follows from our results, this process can be controlled by applying driving current in pulses with the duration $<\Delta t$, which allows one to prevent condensate collapse.

In conclusion, we have demonstrated experimentally a spin-Hall nano-oscillator that enables controllable excitation of the quasi-linear and the bullet dynamical modes. This is facilitated by the injection of spin current into an extended region of the active magnetic film, avoiding the conditions that result in the preferential formation of the bullet mode. Thanks to the ability to excite these fundamentally different modes in the same device, we were able to directly compare their spatial and temporal characteristics, and show that the operation of the SHNOs in the regime of quasi-linear mode oscillations is favorable for the generation of short microwave pulses, while the bullet-mode regime is limited in this respect by the significant time required for the formation of this dynamical state. Our results provide insight into the dynamical mechanisms relevant to practical applications of SHNOs as nano-scale microwave sources.

This work was supported in part by the Deutsche Forschungsgemeinschaft, the NSF Grant No. ECCS 1804198, and Russian Ministry of Science (theme "Spin" No. AAAA-A18-118020290104-2 and project No. 14.Z50.31.0025).

**References**

1. M. I. Dyakonov, V. I. Perel, Sov. Phys. JETP Lett. **13**, 467–469 (1971).

2. J. E. Hirsch, Phys. Rev. Lett. **83**, 1834–1837 (1999).

3. A. Hoffmann, Spin Hall Effects in Metals. IEEE Trans. Magn. **49**, 5172–5193 (2013).

4. J. Sinova, S.O. Valenzuela, J. Wunderlich, C.H. Back, and T. Jungwirth, Rev. Mod. Phys. **87**, 1213 (2015).

5. T. Chen, R. K. Dumas, A. Eklund, P. K. Muduli, A. Houshang, A. A. Awad, P. Dürrenfeld, B. G. Malm, A. Rusu, and J. Åkerman, Proc. of the IEEE **104**, 10 (2016).

6. V. E. Demidov, S. Urazhdin, G. de Loubens, O. Klein, V. Cros, A. Anane, and S. O. Demokritov, Phys. Rep. **673**, 1-31 (2017).

7. L. Liu, C.-F. Pai, D. C. Ralph, and R. A. Buhrman, Phys. Rev. Lett. **109**, 186602 (2012).

8. R. H. Liu, W. L. Lim, and S. Urazhdin, Phys. Rev. Lett. **110**, 147601 (2013).

9. V.E. Demidov, S. Urazhdin, A. Zholud, A.V. Sadovnikov, and S.O. Demokritov, Appl. Phys. Lett. **105**, 172410 (2014).

10. Z. Duan, A. Smith, L. Yang, B. Youngblood, J. Lindner, V. E. Demidov, S. O. Demokritov, and I. N. Krivorotov, Nature Commun. **5**, 5616 (2014).

11. L. Yang, R. Verba, V. Tiberkevich, T. Schneider, A. Smith, Z. Duan, B. Youngblood, K. Lenz, J. Lindner, A. N. Slavin, and I. N. Krivorotov, Sci. Rep. **5**, 16942 (2015).

12. P. Dürrenfeld, A. A. Awad, A. Houshang, R. K. Dumas, and J. Åkerman, Nanoscale **9**, 1285 (2017).

13. A. A. Awad, P. Dürrenfeld, A. Houshang, M. Dvornik, E. Iacocca, R. K. Dumas, and J. Åkerman, Nature Physics **13**, 292–299 (2017).

14. B. Divinskiy, V. E. Demidov, S. Urazhdin, R. Freeman, A. B. Rinkevich, and S. O. Demokritov, Adv. Mater. **30**, 1802837 (2018).

15. M. Collet, X. de Milly, O. d'Allivy Kelly, V. V. Naletov, R. Bernard, P. Bortolotti, J. Ben Youssef, V.E. Demidov, S. O. Demokritov, J. L. Prieto, M. Munoz, V. Cros, A. Anane, G. de Loubens, and O. Klein, Nature Commun. **7**, 10377 (2016).

16. V. E. Demidov, M. Evelt, V. Bessonov, S. O. Demokritov, J. L. Prieto, M. Muñoz, J. Ben Youssef, V.V. Naletov, G. de Loubens, O. Klein, M. Collet, P. Bortolotti, V. Cros, and A. Anane, Sci. Rep. **6**, 32781 (2016).

17. C. Safranski, I. Barsukov, H.K. Lee, T. Schneider, A.A. Jara, A. Smith, H. Chang, K. Lenz, J. Lindner, Y. Tserkovnyak, M. Wu, and I.N. Krivorotov, Nature Commun. **8**, 117 (2017).

18. M. Evelt, L. Soumah, A. B. Rinkevich, S. O. Demokritov, A. Anane, V. Cros, J. B. Youssef, G. de Loubens, O. Klein, P. Bortolotti, and V. E. Demidov, „Emission of coherent propagating magnons by insulator-based spin-orbit torque oscillators," preprint arXiv:1807.09976 (2018).

19. A. N. Slavin and V. S. Tiberkevich, Phys. Rev. Lett. **95**, 237201 (2005).

20. M. Dvornik, A. A. Awad, and J. Åkerman, Phys. Rev. Appl. **9**, 014017 (2018).

21. H. Mazraati, S. R. Etesami, S. A. H. Banuazizi, S. Chung, A. Houshang, A. A. Awad, M. Dvornik, and J. Åkerman, 'Mapping out the spin-wave modes of constriction-based spin Hall nano-oscillator in weak in-plane fields', preprint arXiv:1806.03473 (2018).

22. V. E. Demidov, S. Urazhdin, H. Ulrichs, V. Tiberkevich, A. Slavin, D. Baither, G. Schmitz, and S. O. Demokritov, Nature Mater. **11**, 1028–1031 (2012).

23. M. Ranjbar, P. Dürrenfeld, M. Haidar, E. Iacocca, M. Balinskiy, T. Q. Le, M. Fazlali, A. Houshang, A. A. Awad, R. K. Dumas, and J. Åkerman, IEEE Magn. Lett. 5, 3000504 (2014).

24. V. E. Demidov, H. Ulrichs, S. V. Gurevich, S. O. Demokritov, V. S. Tiberkevich, A. N. Slavin, A. Zholud, and S. Urazhdin, Nature Commun. **2**, 3179 (2014).

25. T. M. Spicer, P. S. Keatley, T. H. J. Loughran, M. Dvornik, A.A. Awad, P. Dürrenfeld, A. Houshang, M. Ranjbar, J. Åkerman, V. V. Kruglyak, and R. J. Hicken, „Time resolved imaging of the non-linear bullet mode within an injection-locked spin Hall nano-oscillator," preprint arXiv:1805.09212 (2018).

26. K. Ando, S. Takahashi, K. Harii, K. Sasage, J. Ieda, S. Maekawa, and E. Saitoh, Phys. Rev. Lett. **101**, 036601 (2008).

27. V. E. Demidov and S. O. Demokritov, IEEE Trans. Mag. **51**, 0800215 (2015).

28. S. M. Rezende, F. M. de Aguiar, and A. Azevedo, Phys. Rev. Lett. **94**, 037202 (2005).

29. A. Slavin and V. Tiberkevich, IEEE Trans. Magn. **45**, 1875-1918 (2009).

30. V. E. Demidov, S. Urazhdin, B. Divinskiy, V. D. Bessonov, A. B. Rinkevich, V.V. Ustinov, and S. O. Demokritov, Nat. Commun. **8**, 1579 (2017).

31. S. O. Demokritov, V. E. Demidov, O. Dzyapko, G. A. Melkov, A. A. Serga, B. Hillebrands, and A. N. Slavin, Nature **443**, 430-433 (2006).

32. H. Ulrichs, V. E. Demidov, and S. O. Demokritov, Appl. Phys. Lett. **104**, 042407 (2014).

33. R. K. Dumas, E. Iacocca, S. Bonetti, S. R. Sani, S. M. Mohseni, A. Eklund, J. Persson, O. Heinonen, and J. Åkerman, Phys. Rev. Lett. **110**, 257202 (2013).

34. S. Bonetti, V. Tiberkevich, G. Consolo, G. Finocchio, P. Muduli, F. Mancoff, A. Slavin, and Johan Åkerman, Phys. Rev. Lett. **105**, 217204 (2010).

**Figure captions**

Fig. 1 (color online) Schematic of the experiment. The inset illustrates the local injection of the electric current and the generation of the pure spin current in the nano-gap between the electrodes.

Fig. 2 (color online) BLS spectra of magnetization oscillations, measured at the labeled values of the driving current; $f_{FMR}$ marks the FMR frequency. The data were obtained at $H_0 = 500$ Oe.

Fig. 3 (color online) (a) Peak intensities and (b) center frequencies of the high-frequency (HF) and the low-frequency (LF) modes vs current $I$; $f_{FMR}$ marks the FMR frequency. (c) Spatial profiles of the HF and the LF modes, measured at $I = 34$ mA. Shadowed area shows the region of the nano-gap. $x=0$ corresponds to the center of the gap. The data were obtained at $H_0 = 500$ Oe.

Fig. 4 (color online) (a) BLS spectra measured at $I = 34$ mA, at different delays with respect to the start of the driving current pulse, as labeled. (b) Temporal evolution of the peak intensities of the auto-oscillation modes, at $I = 34$ mA. Curves show the exponential fits for the leading edge of the time dependence of the dynamic magnetization. (c) Current dependence of the delay time $\Delta t$. Curve is a guide for the eye. The data were obtained at $H_0 = 500$ Oe.

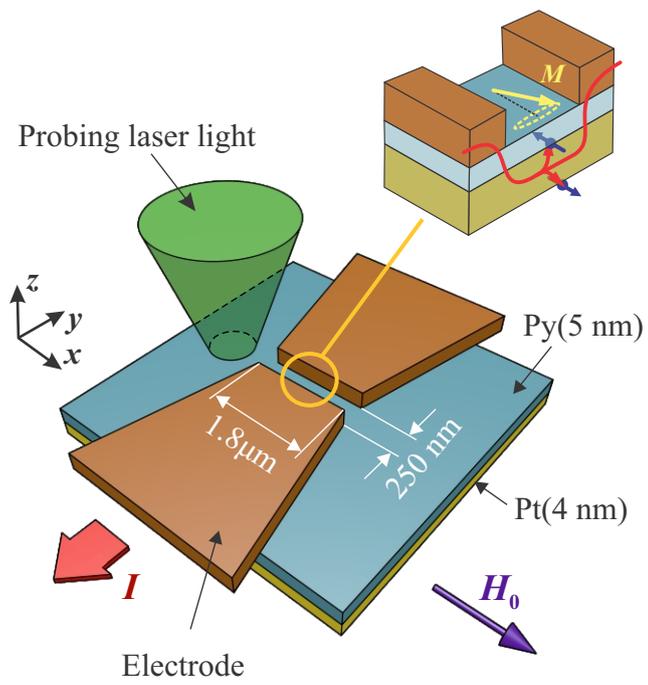

Fig.1

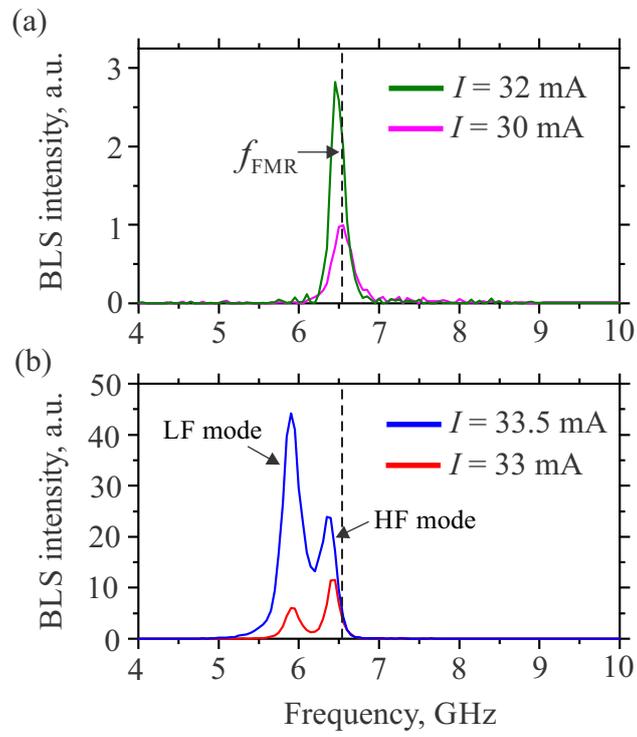

Fig. 2

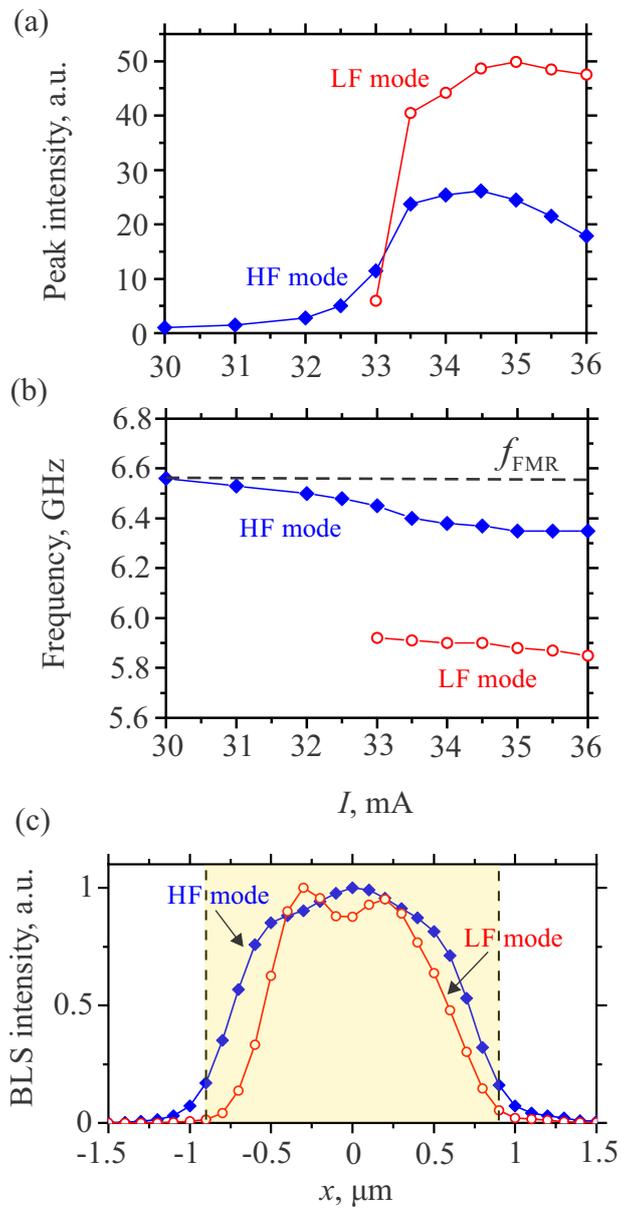

Fig. 3

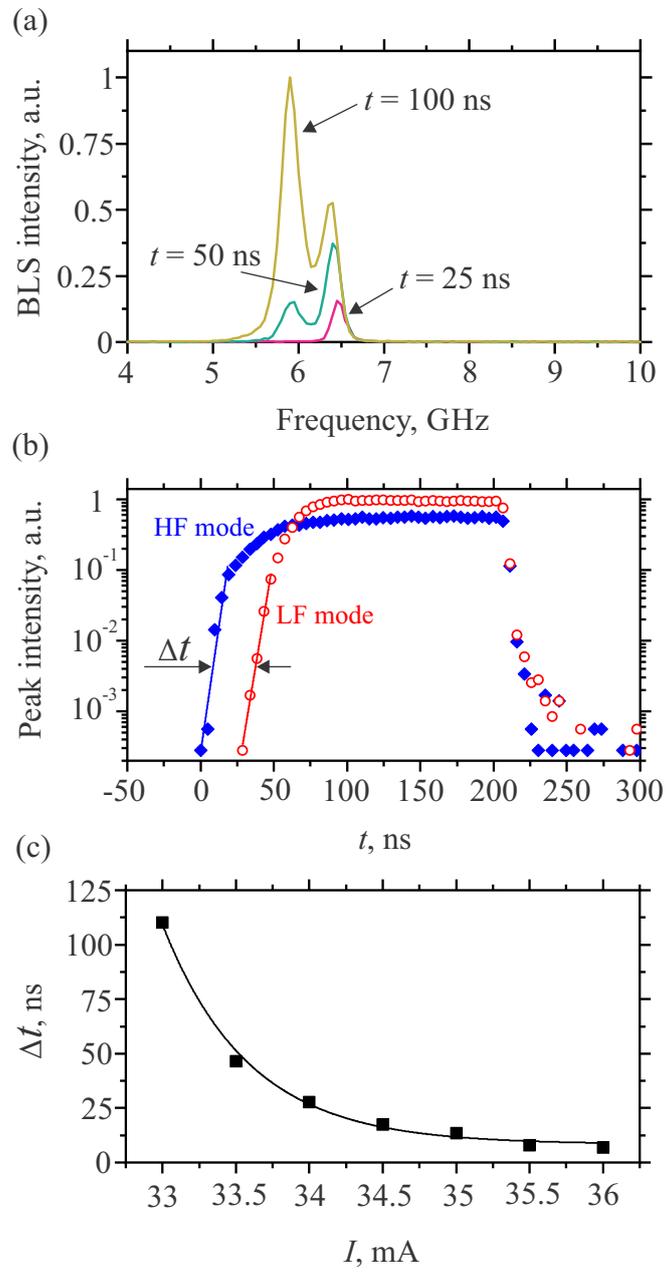

Fig. 4